\documentclass[sigconf,screen]{acmart}
\usepackage{xspace}
\usepackage{tabularx}
\usepackage{booktabs}
\usepackage{multirow}
\usepackage{xcolor}
\usepackage{comment}
\usepackage{amsmath}
\usepackage[ruled,vlined,linesnumbered]{algorithm2e}
\usepackage{verbatim}
\usepackage{tcolorbox}
\usepackage{tabularx}
\usepackage{enumitem}
\usepackage{subcaption}
\usepackage{booktabs}
\usepackage{comment}
\usepackage{microtype}
\usepackage{pbalance}

\usepackage[subtle]{savetrees}

\tcbuselibrary{skins}
\newtcolorbox{rqbox}[1]{
  colback=blue!5!white,
  colframe=blue!30!gray,
  boxrule=0.4pt,
  arc=4pt,
  left=2pt, right=2pt, top=2pt, bottom=2pt
}

\newcommand{\GALA}{\textsf{GALA+}\xspace}
\newcommand{\GALAArxiv}{GALA\xspace}
\newcommand{\STRIX}{\textsf{STRIX}\xspace}
\newcommand{\surescore}{\textsf{SURE-Score}\xspace}
\newcommand{\stitle}[1]{\vspace{0.6mm}\noindent{\bf #1.}}
\newcommand{\GD}{G_\text{D}\xspace}

\AtBeginDocument{%
  }

\setcopyright{cc}
\setcctype{by}
\acmDOI{10.1145/3832783.3834335}
\acmYear{2026}
\copyrightyear{2026}
\acmISBN{979-8-4007-2882-2/2026/10}
\acmConference[ASE '26]{Proceedings of the 41st IEEE/ACM International Conference on Automated Software Engineering}{October 12--16, 2026}{Munich, Germany}
\acmBooktitle{Proceedings of the 41st IEEE/ACM International Conference on Automated Software Engineering (ASE '26), October 12--16, 2026, Munich, Germany}
\acmSubmissionID{ase26main-p46-p}
\received{2026-03-26}
\received[accepted]{2026-06-18}
\begin{document}
\emergencystretch 3em

\title{GALA: Graph-Augmented LLM Agents for Root Cause Analysis and Incident Response in Microservices}

\author{Yifang Tian}
\orcid{0000-0002-8253-2064}
\affiliation{%
  \institution{University of Toronto}
  \city{Toronto}
  \country{Canada}
}
\email{yifang.tian@mail.utoronto.ca}

\author{Yaming Liu}
\orcid{0009-0002-6617-9202}
\affiliation{%
  \institution{University of Toronto}
  \city{Toronto}
  \country{Canada}
}
\email{yaming.liu@mail.utoronto.ca}

\author{Zichun Chong}
\orcid{0009-0000-1020-104X}
\affiliation{%
  \institution{University of Toronto}
  \city{Toronto}
  \country{Canada}
}
\email{zichun.chong@mail.utoronto.ca}

\author{Zihang Huang}
\orcid{0009-0006-4826-7873}
\affiliation{%
  \institution{University of Toronto}
  \city{Toronto}
  \country{Canada}
}
\email{zihang.huang@mail.utoronto.ca}

\author{Yiran Li}
\correspondingauthor
\orcid{0000-0003-4476-016X}
\affiliation{%
  \institution{University of Toronto}
  \city{Toronto}
  \country{Canada}
}
\email{one.li@utoronto.ca}

\author{Hans-Arno Jacobsen}
\orcid{0000-0003-0813-0101}
\affiliation{%
  \institution{University of Toronto}
  \city{Toronto}
  \country{Canada}
}
\email{jacobsen@eecg.toronto.edu}


\begin{abstract}
Microservice root cause analysis (RCA) requires correlating failures across heterogeneous telemetry within complex service dependency graphs. Existing methods often rely on a single telemetry modality; recent LLM-based approaches can suffer from unconstrained exploration and hallucination; and most systems stop at fault ranking without producing actionable incident response. We present \GALA, a graph-augmented LLM agentic framework centered on graph-guided investigation, which uses service dependencies to bound exploration and refine diagnosis through localized multi-modal evidence. For initial hypothesis generation, \GALA combines complementary telemetry signals with \STRIX, a novel trace- and graph-structure-aware scoring module. \GALA then produces ranked diagnoses, incident summaries, and stratified action recommendations. We further introduce \surescore, a human-guided evaluation framework co-developed with industry SRE experts for assessing RCA-specific output quality beyond conventional text similarity metrics. On two microservice benchmarks, \GALA consistently achieves the strongest overall results, surpassing the best LLM-based baseline by more than 25 percentage points in AC@1, while also receiving the highest ratings from both \surescore and independent human SRE evaluation.

\end{abstract}




\maketitle

\section{Introduction}

Modern software systems increasingly adopt microservice architectures for scalability, maintainability, and deployment flexibility~\cite{dragoni2017microservices}. However, decomposing applications into numerous loosely coupled services communicating through intricate dependency graphs introduces significant operational complexity. When failures occur, root cause analysis (RCA) requires correlating symptoms across multiple services, analyzing heterogeneous telemetry data, and understanding complex inter-service dependencies---all under tight time pressure. Beyond identifying the faulty service, site reliability engineers (SREs) need a concise explanation of the failure's propagation path and actionable steps to resolve the incident~\cite{zhang2025failure}.

\begin{figure}[h]
    \centering
    \centerline{
        \includegraphics[width=0.47\textwidth]{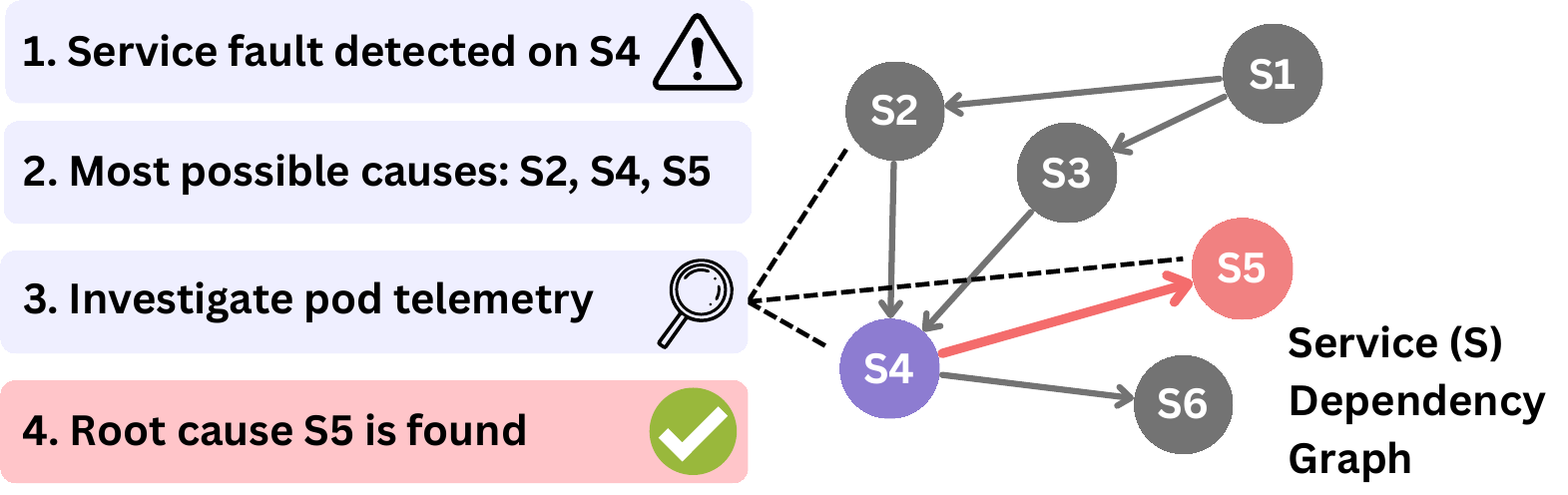}
    }
    \vspace{-2mm}
    \caption{Motivating example: a fault detected on $S_4$ is traced through the dependency graph to the true root cause $S_5$.}
    \label{fig:motivating_example}
\end{figure}

Consider a cloud-native application with six microservices $S_1$--$S_6$ and the dependency graph in Figure~\ref{fig:motivating_example}. In one incident, an alert is triggered on $S_4$ after its latency exceeds the acceptable threshold. An engineer’s initial inspection narrows the likely sources to $S_4$ itself, its upstream caller $S_2$, and its downstream callee $S_5$. However, the fault does not originate at the alerted service $S_4$: while $S_4$ shows elevated latency, the underlying issue is a CPU-intensive slowdown in $S_5$, whose delayed responses propagate upstream and inflate $S_4$’s latency. This example illustrates three core challenges. First, the service where the problem is observed may differ from the true fault origin, so RCA must reason over service dependencies rather than inspect services in isolation. Second, different telemetry signals may point to different suspects, so accurate diagnosis requires combining multiple modalities rather than relying on any single one. Third, identifying the faulty service alone is insufficient; engineers also need an explanation of how the failure propagated and what actions should be taken to mitigate it. In this work, we address all three challenges.

However, existing approaches still fall short of addressing these three challenges in a unified manner. Statistical methods~\cite{lin2024root,pham2024baro,xin2023causalrca} address only challenge~(2) partially: they operate on single modalities or use static heuristics that miss dynamic failure modes such as memory leaks triggered by specific API sequences~\cite{xie2024cloud}. Multi-modal approaches~\cite{fu2025msfsanomaly,yu2023nezha,zheng2024multi} fuse heterogeneous data but collapse it into unified representations, potentially losing the modality-specific signals that distinguish propagation victims from true root causes. 
Recent large language model (LLM) advancements have opened new possibilities for automating complex reasoning~\cite{brown2020language}, with chain-of-thought~\cite{wei2022chain}, tree-of-thought~\cite{yao2023tree}, and agentic workflows such as ReAct~\cite{yao2023react} demonstrating potential for multi-step problem-solving. However, our empirical evaluation shows that unconstrained multi-agent reasoning may search across the entire service space without topological grounding, incurring excessive search overhead, hallucination, and hypothesis drift that ultimately degrades precision. Across all paradigms, most methods stop at ranking potential root causes, addressing neither challenge~(3) nor the evaluation gap it creates: standard natural language generation (NLG) metrics (BLEU~\cite{papineni2002bleu}, ROUGE~\cite{lin2004rouge}, BERTScore~\cite{zhang2020bertscore}) measure semantic overlap but cannot assess the causal grounding and operational specificity that SREs require. Effective LLM-based RCA therefore demands topology-constrained reasoning that bounds exploration to causally plausible paths, maintains cross-modal consistency, and produces not just fault rankings but also structured incident response.

We present \GALA (\underline{G}raph-\underline{A}ugmented \underline{L}LM \underline{A}gentic Workflow), a framework that addresses these limitations by constraining agentic reasoning steps to the service dependency graph. To overcome single-modality blindness, \GALA combines metrics-based causal ranking with \STRIX, a novel scoring module that exploits trace patterns and dependency-graph topology, and consolidates their complementary signals into a unified candidate set. Each candidate's raw telemetry is distilled into a compact diagnostic bundle preserving temporal patterns, dependency context, and error signals.  To avoid the unfocused exploration that can arise in unconstrained agentic reasoning, we propose a graph-guided investigation mechanism in which parallel agents follow fault-propagation edges with confidence-driven, depth-bounded expansion, keeping the search localized to a small neighborhood around plausible root-cause candidates. The resulting evidence is synthesized into a final ranking, structured incident summaries, and stratified action recommendations. We further introduce \surescore, a human-guided LLM evaluation framework co-developed with industry SRE experts for RCA-specific output assessment.

Evaluated on two microservice benchmarks, \GALA achieves 74.44\% AC@1 on OnlineBoutique and 73.33\% on TrainTicket, outperforming the strongest baseline by over 25\%, with the highest \surescore ratings across all dimensions confirmed by both dual-LLM grading and independent human SRE assessment. Our contributions are threefold:

\begin{enumerate}[leftmargin=*]

\item \textbf{Dual-Signal Root Cause Hypothesis Generation}: We develop a hypothesis generation mechanism that fuses metrics-based initial ranking with \STRIX, a novel trace- and graph-structure-aware scoring module, to produce complementary signals consolidated into a unified ranked candidate set.

\item \textbf{Graph-Guided Agentic Investigation}: We introduce a topology-constrained parallel investigation framework in which agents follow fault-propagation edges with confidence-driven, depth-bounded expansion, substantially reducing the search space while preserving diagnostic completeness.

\item \textbf{Incident Response Generation and Evaluation}: We build a pipeline that produces ranked diagnoses, structured incident summaries, and stratified action recommendations. These outputs are evaluated with \surescore, a human-guided framework on which \GALA achieves the highest scores across all four dimensions in both automated and human assessment.

\end{enumerate}

\section{Problem Statement}
\label{sec:problem-statement}

\begin{figure*}[t]
    \centering
    \centerline{
        \includegraphics[width=\textwidth]{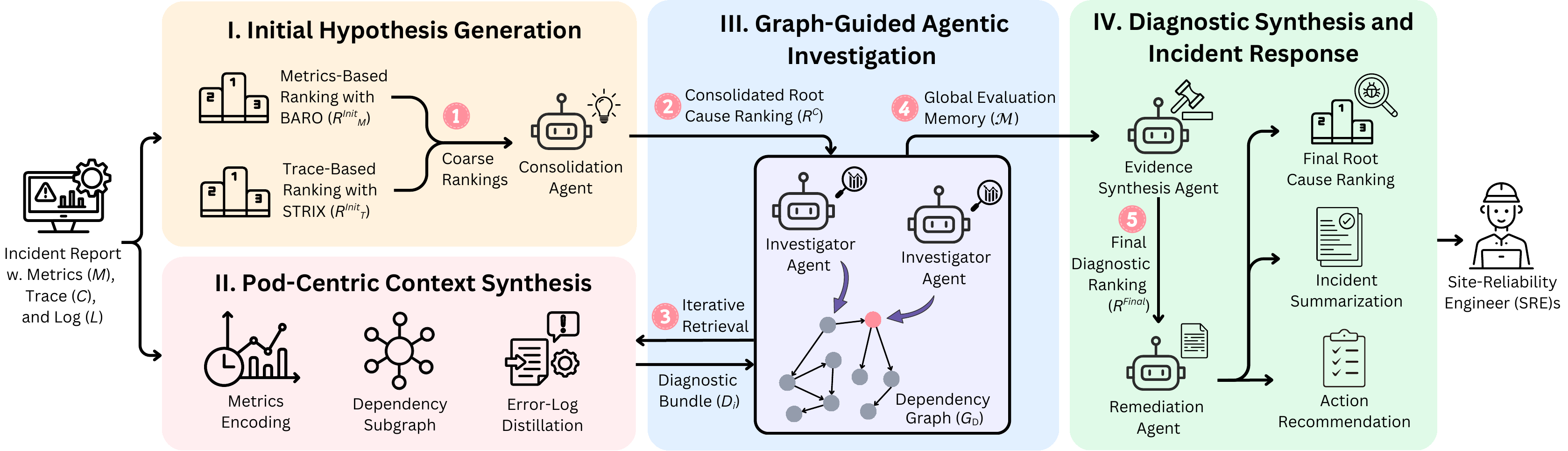}
    }
    \vspace{-2mm}

    \caption{Overview of \GALA.
    }

    \label{fig:GALA}
\end{figure*}

We consider microservice architectures~\cite{dragoni2017microservices} where a \textit{pod} is the smallest deployable unit. Telemetry comprises three modalities: \textit{metrics} (quantitative measurements such as CPU usage and latency~\cite{bhosale2022metrics}), \textit{logs} (timestamped event records~\cite{yu2023logreducer}), and \textit{traces} (cross-service call relationships and timing). In distributed tracing, a span represents a single operation within a service, identified by a \texttt{spanID} and associated with a global \texttt{traceID}. Spans reference parent spans via \texttt{parentSpanID}, enabling construction of a \textit{service dependency graph}: a directed acyclic graph (DAG) where nodes represent services and directed edges encode call relationships~\cite{sigelman2010dapper}.

Let $\mathcal{P}$ denote the set of pods in a microservice deployment experiencing an incident. For each pod $i \in \mathcal{P}$, we observe metrics $m_i$, logs $l_i$, and traces $t_i$ collected during the incident window, with $M$, $L$, $T$ denoting the complete collections across all pods. A root cause is defined as the specific pod and failure mode (e.g., CPU hog in \texttt{order-service}) that, when addressed, resolves the incident. 

Given $\{M, L, T\}$, the RCA task requires producing:
\begin{enumerate}[leftmargin=*]
    \item \textbf{Ranked Root Cause Identification}: A prioritized ranking of candidate pods by fault likelihood.
    \item \textbf{Incident Summary}: A human-readable narrative explaining the symptom-to-root-cause causal chain across all three telemetry modalities.
    \item \textbf{Prioritized Action Recommendations}: Three remediation steps stratified across temporal horizons, covering immediate mitigation, permanent fix, and preventative measure.
\end{enumerate}

\section{The \GALA Design} 
\label{sec:methodology}

In this section, we present \GALA, a unified multi-modal framework for root-cause analysis in microservice-based systems. As shown in Figure~\ref{fig:GALA}, \GALA operates in four coordinated phases that progressively transform raw telemetry into root cause insight and actionable remediation guidance. Phase~I (\textit{Initial Hypothesis Generation}) analyzes system-wide metrics and traces to produce a consolidated initial ranking of likely root-cause pods. Phase~II (\textit{Pod-Centric Context Synthesis}) distills each candidate pod’s metrics, logs, and traces into a compact context that preserves its temporal pattern, dependency neighborhood, and error signals. Phase~III (\textit{Graph-Guided Agentic Reasoning and Re-ranking}) dispatches investigator agents to examine high-ranked pods and traverse their service dependency graph, using localized evidence to iteratively refine the ranking. Phase~IV (\textit{Final Output Preparation}) consolidates these agentic evaluations into a final root-cause ranking and a structured incident summary with prioritized remediation recommendations.

\subsection{Initial Hypothesis Generation}
\label{Initial Hypothesis Generation}
Before dispatching LLM agents for deep pod‑level investigation, \GALA first requires a coarse estimate of which services are most likely responsible for the incident. 
In Phase I, this initial hypothesis is provided efficiently by analyzing metrics and traces with modality‑specific methods, including our novel trace analysis technique \STRIX. The resulting modality‑level rankings are then integrated into a reliable preliminary root‑cause estimate.
Logs are excluded at this stage because many faults manifest silently without any error entries; their semantic content is instead analyzed in later phases.

\subsubsection{Coarse Metrics-based Ranking}

To generate an initial hypothesis from metrics, \GALA instantiates this stage with BARO~\cite{pham2024baro}, which first identifies anomalous services via Bayesian online change-point detection and then constructs a causal DAG $G_M$ over the collected metrics data $M$ to model inter-service dependencies and failure propagation paths. 
The resulting DAG is then used directly to score each candidate service by its causal influence on the observed anomalies, yielding an initial ranked list $R^{\mathrm{Init}}_M = \text{BARO}(M)$.

Nevertheless, metrics-based causal discovery methods, such as BARO, have two limitations reported in the literature~\cite{pham2024root}: they often misidentify causal edge directions, and inherently cannot distinguish genuine anomalies from routine operational fluctuations. To mitigate both shortcomings, we introduce a complementary trace‑based ranking that directly examines dependency directions and performance stability.

\subsubsection{Coarse Trace-based Ranking}
\label{sec:strix}

As an orthogonal module to the metrics‑based causal inference, we introduce \STRIX (\underline{S}tructural \underline{T}race \underline{R}anking via \underline{I}ntermodal e\underline{X}ploration), which derives an initial root cause ranking from distributed traces. As a topology‑driven method, \STRIX analyzes the service dependency graph and inspects span‑level latency patterns to surface a complementary hypothesis. Formally, given trace corpus $T$, \STRIX produces an initial ranked list $R^{\mathrm{Init}}_{T} = \text{STRIX}(T)$.

\stitle{Service Dependency Graph Construction} \STRIX first constructs a weighted directed dependency graph $\GD = (V, E)$ from the collected trace corpus $T$, where $D$ denotes the service dependency. Each node $v \in V$ represents a microservice, and each directed edge $(u, v) \in E$ represents the aggregated invocation relationship from service $u$ to service $v$, consolidated across all observed spans in $T$. Edge and node attributes are populated by aggregating span-level measurements across all matching records, including latency statistics (mean, P50, P95, P99, max, standard deviation, and total cumulative latency) 
and call volume (i.e., the total number of spans observed on that edge or node). These attributes form the basis for the three diagnostic dimensions computed in the next step. 

\stitle{Three Diagnostic Dimensions} Based on this service dependency graph, \STRIX evaluates three diagnostic dimensions for each service node $v$, integrating both the graph’s topology and its aggregated attribute statistics.
\begin{itemize}[leftmargin=*]
    \item \textbf{Instability} ($\phi_1$) captures the hallmark of resource contention and transient faults: a service under stress develops a pronounced latency tail while its median remains comparatively stable~\cite{sigelman2010dapper}. This 
    separates true bottlenecks exhibiting sharp latency spikes from services that are simply slow but predictable. We define
    $\phi_1(v) = \frac{p_{99}(v)}{p_{50}(v)},$
    the ratio of tail latency ($p_{99}$) of incoming calls to the median  ($p_{50}$) as the Instability score for service $v$.

    \item \textbf{Centrality} ($\phi_2$) captures the structural tendency that services with high fan-in and strong graph centrality are more likely to be true root causes, whereas high fan-out services are often fault-propagation intermediaries~\cite{wu2020microrca}.
    We formalize this by weighting PageRank with the fan-in to fan-out ratio:
    $$\phi_2(v) = \text{PageRank}(v) \cdot \frac{1 + \text{in-degree}(v)}{1 + \text{out-degree}(v)}.$$
    This jointly measures global structural importance via PageRank and local dependency pattern via the degree ratio, elevating services that are widely reachable and heavily depended upon.

    \item \textbf{Impact} ($\phi_3$) captures how much delay originates from a service’s own processing rather than from waiting on downstream callees, a decomposition established in distributed trace analysis~\cite{li2021practical}. A service whose average latency substantially exceeds that of its slowest callee is more likely to be the true bottleneck. Let $\text{AvgLatency}(v)$ denote the mean span duration of service $v$ across all observed calls, and $\text{TotalLatency}(v)$ its cumulative span duration aggregated over all spans. We define
$$\phi_3(v) = \text{TotalLatency}(v) \cdot \max\!\left(0,\, 1 - \frac{\max_{(v,u) \in E}\, \text{AvgLatency}(u)}{\text{AvgLatency}(v)}\right),$$
where the bracketed term estimates the fraction of $v$'s average latency attributable to its own processing by discounting the maximum average latency of its outgoing callees, and $\text{TotalLatency}(v)$ scales this local inefficiency by the volume of affected traffic to produce a globally comparable impact score.

\end{itemize}

\stitle{Rank Aggregation via Reciprocal Rank Fusion} Since $\phi_1$, $\phi_2$, and $\phi_3$ span fundamentally different numerical ranges, direct weighted combination would require domain-specific calibration. \STRIX instead applies Reciprocal Rank Fusion (RRF)~\cite{cormack2009reciprocal}, which operates on ordinal rank positions rather than raw scores and is thus scale-invariant by construction. Each service $v$ receives an RRF score aggregated across all three dimensions:
$\text{RRF}(v) = \sum_{j=1}^{3} \frac{1}{\epsilon + \text{rank}(v, \phi_j)}$
where $\epsilon=60$ is a smoothing constant following the original implementation~\cite{cormack2009reciprocal}. With RRF, services that rank consistently high across all three dimensions are naturally elevated, while those that dominate in only one are moderated.

The final output of \STRIX is the ranked list $R^{\text{Init}}_T = \text{STRIX}(T)$, defined as the services ordered by their RRF scores in descending order. This provides a robust, topology-grounded initial hypothesis that is structurally complementary to the time-series-driven $R^{\text{Init}}_M$, and the two are subsequently consolidated in Section~\ref{sec:hypothesis_fusion}.

\subsubsection{Consolidation Agent and Hypothesis Fusion} 
\label{sec:hypothesis_fusion}

The two coarse rankings above carry complementary but asymmetric information. $R^{\text{Init}}_M$ provides service-fault-type pairs grounded in causal structure learning, while $R^{\text{Init}}_T$ provides topology-driven service orderings without fault-type attribution. A candidate appearing only in $R^{\text{Init}}_T$ (hereafter \emph{trace-only}) may be a structurally prominent bottleneck whose fault type is unknown, whereas a candidate appearing only in $R^{\text{Init}}_M$ (\emph{metrics-only}) may carry a precise fault label but lack corroborating structural evidence. Naive score-based merging would ignore this asymmetry, so we design a \textsf{Consolidation Agent} that reasons explicitly over each case to produce a unified ranking.

Given $R^{\text{Init}}_M$ and $R^{\text{Init}}_T$ as input, the agent is prompted to: (1) identify cross-modal consensus candidates that appear prominently in both rankings and assign them elevated priority; (2) infer plausible fault types for trace-only candidates by leveraging fault-type context from $R^{\text{Init}}_M$; and (3) retain metrics-only candidates with appropriately reduced confidence when trace evidence is absent. Each candidate in the output is assigned a consolidated confidence score $\hat{c}_i \in [0,1]$ with explicit evidence provenance drawn from one or both modalities, reflecting the strength of cross-modal corroboration: candidates confirmed by both rankings receive the highest $\hat{c}_i$, trace-only candidates receive moderate scores with inferred fault types, and metrics-only candidates receive the lowest. This provenance-aware scoring ensures that downstream agents can distinguish well-substantiated hypotheses from speculative ones. The result is a fused ranking $R^{\text{C}} = \{(v_i, \hat{c}_i)\}$ whose
top-$k$ candidates, each paired with its consolidated confidence, are forwarded to the graph-guided investigation in Section~\ref{LLM Agentic Reasoning and Re‐ranking}.

\subsection{Pod-Centric Context Synthesis}
\label{Pod-Centric Context Synthesis}
Raw metrics, logs, and traces are too voluminous and heterogeneous to be passed directly to an LLM. Therefore, in Phase II, each candidate pod’s telemetry is condensed into a compact diagnostic bundle $D_i$ that preserves the key temporal patterns, dependency context, and error signals needed for targeted pod-level analysis.

For each pod $i$ and each monitoring metric, we serialize per-second measurements within the incident window $[\tau - \Delta\tau,\, \tau + \Delta\tau]$ into a timestamped JSON dictionary $\mathbf{m}_i^{*}$, where $\tau$ is the anomaly onset timestamp derived from the incident alert and $\Delta\tau$ is a fixed look-around window~\cite{zhang2025failure}. This representation replaces an earlier Base64-encoded line-chart design; empirically, JSON reduces token consumption while improving diagnostic accuracy. For structural context, we extract a 1-hop subgraph $g_i$ from the service dependency graph, comprising $v_i$, its direct predecessors and successors, and associated edge attributes such as call volume and latency statistics. For log evidence, a \textsf{LogDistiller} processes the raw stream $l_i$ by retaining only error-level and exception entries, ordered by timestamp and deduplicated at the message level. If the retained entries exceed a token budget, we keep a representative sample; otherwise, we supplement them with a small number of non-error entries to distinguish silent failures from log absence, yielding $l_i^{*}$.

These three artifacts are then assembled into the diagnostic bundle $D_i = \{\mathbf{m}_i^{*},\, g_i,\, l_i^{*}\}$ using a structured text template with three labeled sections: $\mathbf{m}_i^{*}$ as JSON, $g_i$ as an annotated adjacency list, and $l_i^{*}$ as a chronologically ordered message list. This format preserves each modality in its native representation within a single token-efficient context.

\subsection{Graph-Guided Agentic Investigation}
\label{LLM Agentic Reasoning and Re‐ranking}

Given the initial ranking $R^{\text{C}}$, Phase III performs a graph-guided investigation that re-examines highly ranked hypothesis candidates through localized multi-modal reasoning and extends to neighboring services when local evidence is insufficient. The key idea is to couple LLM-based agentic analysis with the structure of the service dependency graph: each \textsf{Investigator Agent} starts from a candidate in $R^{\text{C}}$, reasons over its diagnostic bundle $D_i$, and, when the evidence is inconclusive, examines services in its local dependency neighborhood. The structured assessments produced along these investigation branches are accumulated in a shared investigation memory $\mathcal{M}$ for downstream synthesis. In this way, \GALA deepens the investigation beyond the initial ranking while keeping exploration confined to a small, structurally meaningful neighborhood around plausible root-cause candidates.

\begin{algorithm}[t]
\caption{Graph-Guided Investigation}
\label{alg:parallel_investigation}
\small
\SetKwFunction{FInvestigate}{DeepDive}
\SetKwProg{Fn}{Function}{:}{}
\KwIn{Consolidated ranking $R^{\text{C}}$, service dependency graph $G{=}(V,E)$,
diagnostic bundles $\{D_i\}$, parallel branches $k$, threshold $\theta$,
batch size $b$}
\KwOut{Global evaluation memory $\mathcal{M}$}

$\mathcal{M} \leftarrow \varnothing$\;
\For{each $(v_i, \hat{c}_i) \in \mathrm{TopK}(R^{\text{C}},\ k)$ \textbf{in parallel}}{
    \FInvestigate({$v_i$})\;
}

\Fn{\FInvestigate{$v$}}{
    $\hat{c}_v, \sigma_v^{\text{evid}}, \sigma_v^{\text{causal}}, F_v^+, F_v^-
        \leftarrow \mathrm{Investigator}(D_v,\ R^{\text{C}})$\;
    $\mathcal{M} \gets \mathcal{M} \cup
        \{(v,\ \hat{c}_v,\ \sigma_v^{\text{evid}},\ \sigma_v^{\text{causal}},\
        F_v^+,\ F_v^-)\}$\;
    \If{$\hat{c}_v \geq \theta$}{
        \Return\tcp*[r]{Sufficient local evidence}
    }
    $\mathcal{N}(v) \leftarrow
        \mathrm{Predecessors}(v,G)\ \cup\ \mathrm{Successors}(v,G)$\;
    $\mathcal{U} \leftarrow \mathcal{N}(v)$\;

\While{$\mathcal{U} \neq \varnothing$}{
    $\mathcal{B} \leftarrow
        \mathrm{RandomSample}\!\left(
            \mathcal{U},\ \min\{b,|\mathcal{U}|\}
        \right)$\;
    $\mathcal{U} \leftarrow \mathcal{U} \setminus \mathcal{B}$\;

    \For{each $u \in \mathcal{B}$}{
        $\hat{c}_u, \sigma_u^{\text{evid}}, \sigma_u^{\text{causal}},
        F_u^+, F_u^-
        \leftarrow
        \mathrm{Investigator}
        (D_u,\ R^{\text{C}})$\;

        $\mathcal{M} \leftarrow \mathcal{M} \cup
        \{(u,\ \hat{c}_u,\ \sigma_u^{\text{evid}},
        \sigma_u^{\text{causal}},\ F_u^+,\ F_u^-)\}$\;
    }

    \If{$\max_{u\in\mathcal{B}}\hat{c}_u \geq \theta$}{
        \Return\tcp*[r]{High-confidence neighbor}
    }
}
}
\end{algorithm}

The Graph-Guided Investigation Algorithm (Algorithm~\ref{alg:parallel_investigation}) formalizes this process. Starting from the top-$k$ entries of $R^{\text{C}}$, \GALA invokes \textsc{DeepDive} on each candidate. For a service $v$, the \textsf{Investigator Agent} reasons over the diagnostic bundle $D_v$ together with the consolidated ranking $R^{\text{C}}$, and produces a structured assessment consisting of a root-cause confidence $\hat{c}_v \in [0,1]$, an evidence summary $\sigma_v^{\text{evid}}$, a causal explanation $\sigma_v^{\text{causal}}$, and supporting and contradicting factors $F_v^+$ and $F_v^-$. Every assessment is committed to the shared memory $\mathcal{M}$ as soon as the agent returns, so that $\mathcal{M}$ retains a record of every investigated service.

The resulting confidence determines whether the investigation stops or extends. If $\hat{c}_v \geq \theta$, the candidate's local evidence is considered sufficient and the branch terminates. Otherwise, \GALA examines the candidate's dependency neighbors in randomized batches of size $b$, stopping as soon as a batch contains a high-confidence service or all neighbors have been examined.

The investigation is bounded by the size of each candidate's local neighborhood. A candidate with sufficient local evidence requires one assessment; otherwise, at most $|\mathcal{N}(v)|$ additional services are examined. Across the top-$k$ candidates, the worst-case work is $O(k\Delta)$, where $\Delta$ is the maximum degree of $G$, while the $k$ branches execute concurrently.

This design gives \GALA a bounded search procedure that adapts to the strength of local evidence. Rather than relying solely on the initial ranking or allowing unconstrained agentic exploration, it performs additional reasoning only when a candidate's local evidence is inconclusive, and restricts that reasoning to structurally related services. After all investigations terminate, $\mathcal{M}$ contains the structured assessments for the explored services, which are then passed to the \textsf{Evidence Synthesis Agent} in Phase IV for final ranking and incident-report generation.

\subsection{Diagnostic Synthesis and Incident Response}
\label{Final Output Preparation}
In Phase IV, branch-specific findings collected during graph-guided investigation are transformed into a single coherent diagnosis and an evidence-grounded incident response. Its key role is not merely to summarize the contents of investigation memory $\mathcal{M}$, but to reconcile potentially competing agent assessments, preserve consistency with the initial system-level signals, and convert the final diagnosis into remediation guidance that is directly tied to the inferred failure mechanism. Through this design, \GALA bridges the gap between root-cause identification and operational incident handling.

\stitle{Evidence Synthesis}
Once all \textsf{Investigator Agents} terminate, the \textsf{Evidence Synthesis Agent} synthesizes the contents of $\mathcal{M}$ into a final diagnostic ranking $R^{\text{Final}}$. Rather than relying on LLM reasoning alone, this synthesis is governed by two arbitration rules. First, \textit{signal anchoring} treats the initial confidence scores from $R^{\text{C}}$ as grounded priors, preventing hallucination-driven re-ranking in which a locally persuasive assessment overrides stronger system-level evidence. Second, \textit{chain coherence} requires that the top-ranked candidate’s causal reasoning chain $\sigma_v^{\text{causal}}$ explain the downstream symptoms observed in other explored branches through their evidence summaries $\sigma_u^{\text{evid}}$ and contradicting factors $F_u^-$. A candidate is promoted to rank 1 only if it not only appears locally plausible, but also serves as the most coherent global explanation of the evidence distributed across $\mathcal{M}$. This synthesis step is critical for distinguishing true root causes from downstream victims that may exhibit strong local symptoms.

\stitle{Remediation Consolidation}
Building on $R^{\text{Final}}$ and its associated reasoning traces, the \textsf{Remediation Agent} generates an incident report that couples diagnosis with evidence-grounded action recommendations. Specifically, it takes as input the final ranking, the supporting rationale, and the inferred causal chain of failure propagation, and produces a structured incident summary together with three remediation actions stratified across temporal horizons: an \textit{immediate mitigation} to restore service stability, a \textit{permanent fix} that addresses the underlying failure mechanism, and a \textit{preventative measure} tailored to the observed propagation pattern and dependency topology traversed in Phase III. The resulting recommendations are incident-specific, traceable to the synthesized evidence, and directly actionable for SREs.

Taken together, Phase IV produces a final root-cause ranking, a structured incident summary, and prioritized remediation recommendations spanning immediate mitigation, permanent fix, and future prevention. By converting investigation findings into a coherent diagnosis and actionable response plan, \GALA helps SREs move from root-cause identification to incident resolution more quickly and with stronger evidential support.

\section{Human-Guided Evaluation of RCA Reports}
\label{SURE_Score}
\begin{figure}[h]
    \centering
    \centerline{
        \includegraphics[width=0.48\textwidth]{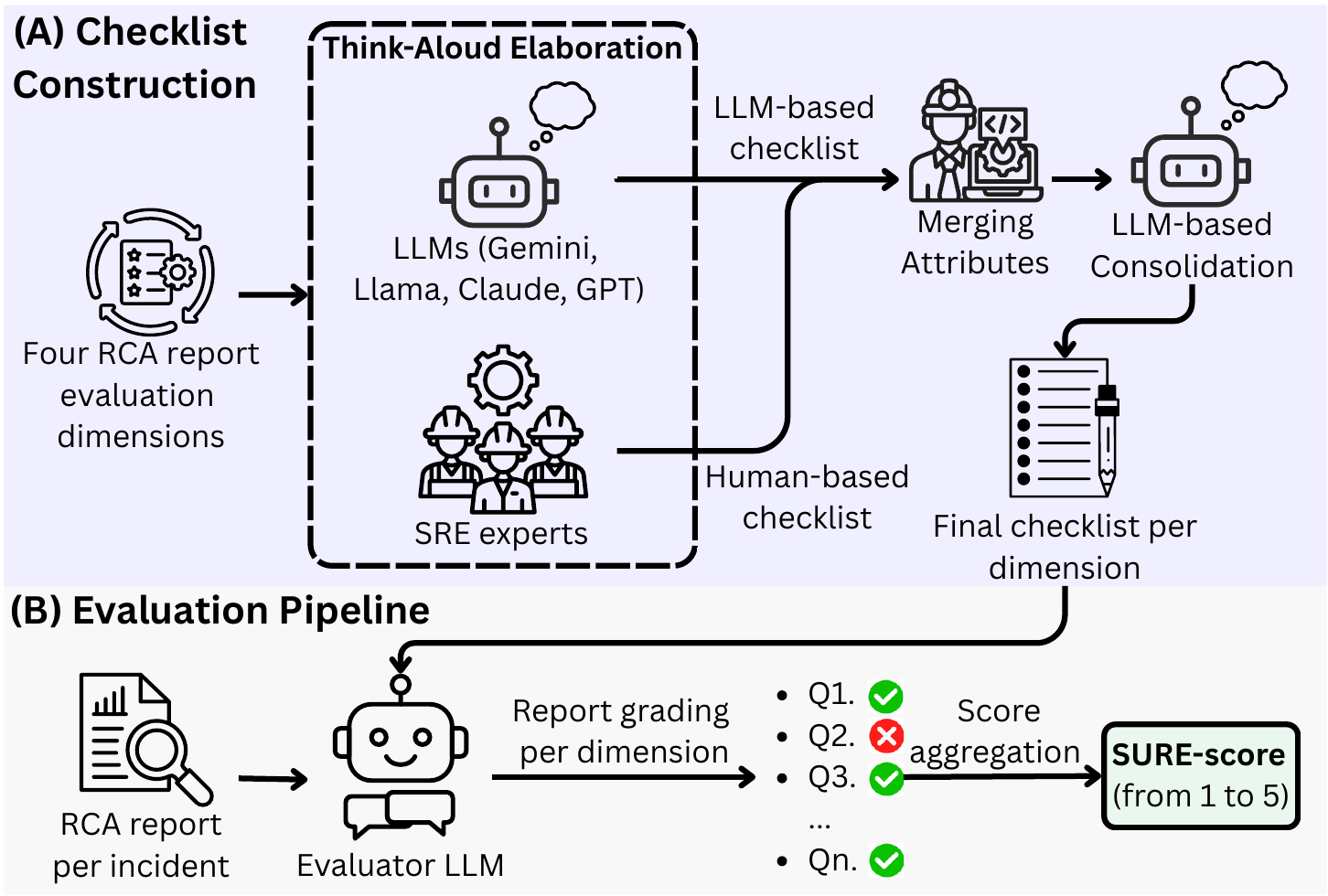}
    }
    \vspace{-3mm}
    \caption{SURE-Score framework.} 
    \label{fig:SURE_Score_pipeline}
\end{figure}
Evaluating RCA-generated summaries and remediation recommendations should focus on whether the outputs are logically coherent, causally grounded in the observed failure, and operationally useful for incident response. However, standard natural language generation metrics such as BLEU~\cite{papineni2002bleu}, ROUGE~\cite{lin2004rouge}, and BERTScore~\cite{zhang2020bertscore} primarily capture lexical or semantic overlap and therefore fail to assess these RCA-specific dimensions. In contrast, LLM-based evaluation frameworks that incorporate human guidance and feedback can provide more reliable and contextually valid assessments than such automated metrics~\cite{chu2025thinkworkbettercombining}. To address this gap, we introduce \surescore\ (\underline{SU}mmarization \underline{RE}commendation Score), a novel human-guided evaluation framework for RCA-specific textual outputs, organized into two steps (Figure~\ref{fig:SURE_Score_pipeline}).

\stitle{Step A: Checklist Construction} \surescore adopts the Think-Aloud Elaboration protocol from InteracEval~\cite{chu2025thinkworkbettercombining}, which provides a mechanism for both human experts and LLMs to externalize their reasoning about what constitutes a high‑quality RCA report. Together with domain experts at Company A, we first co‑develop four RCA‑critical evaluation dimensions:

\begin{enumerate} [leftmargin=*]

\item \textbf{Evidence Groundedness}, which requires every reasoning step in the narrative to be explicitly supported by concrete diagnostic evidence, such as a log entry, metric deviation, or trace anomaly, while penalizing hallucinations, unsupported claims, and steps that cannot be traced to observable artifacts.

\item \textbf{Operational Feasibility}, which evaluates whether the proposed remediation actions are concrete and implementable within the system architecture, severely penalizing vague directives such as ``improve monitoring''.

\item \textbf{Diagnostic Precision}, which requires specific service names, pod identifiers, and quantitative impact metrics (e.g., "latency increased from 200ms to 1.5s") rather than generic placeholder descriptions.

\item \textbf{Triage Logic}, which evaluates whether the summary follows a professional triage flow from symptoms to pivot evidence to root cause, with critical findings presented first and a coherent reasoning pathway synthesized for the SREs.
\end{enumerate}

Eight SRE experts from Company A, with four to ten years of operations experience, and four LLMs (\textit{GPT-4.1-mini}~\cite{openai_gpt41_mini_2025}, \textit{Gemini-2.5-Pro}~\cite{gemini_2_5_pro_2025}, \textit{Claude-Sonnet-4.5}~\cite{claude_sonnet_4_5_2025}, and \textit{Llama-4-Maverick}~\cite{Llama-4-Maverick_2025}) then individually perform Think‑Aloud elaboration for each dimension, generating candidate attributes that reflect their understanding of what strong RCA performance entails. We use \textit{Qwen3}~\cite{Qwen3_80B_2025} to merge and de-duplicate the collected attribute lists into a final, dimension-specific binary checklist of Yes/No criteria. This multi-source construction ensures that the checklist integrates practitioners’ operational intuitions with LLM-derived perspectives, yielding an evaluation instrument that is both practically grounded and systematically structured.

\stitle{Step B: Evaluation Pipeline}
We first construct a validated reference report for each failure case: we prompt an LLM with the ground-truth faulty pod, failure mode, and fault onset time, together with a description of the microservice system and SRE-authored few-shot examples, and constrain the output to a fixed template comprising an incident summary and three stratified remediation recommendations. SRE experts then manually verify each reference, confirming that the diagnosis is consistent with the ground-truth fault type and that the recommended actions are operationally plausible in the target system.
For each RCA report, \textit{Qwen3}~\cite{Qwen3_80B_2025} and \textit{Llama-4-Maverick}~\cite{Llama-4-Maverick_2025} independently evaluate the report against the constructed checklist, using the validated reference as ground truth. 
Each binary criterion is assessed separately, and the positive responses are aggregated within each dimension and rescaled to a 1--5 scale, yielding an overall \surescore\ from each evaluator. To reduce self-preference bias~\cite{chen2025beyond}, these evaluator models are chosen to be different from the LLM used to generate the report under assessment. This cross-model grading design, together with the human-validated checklist, mitigates the risk that an evaluator favors its own generation style. We adopt this framework following InteracEval, which has been empirically validated to align well with human judgments~\cite{chu2025thinkworkbettercombining}.

\section{Evaluation}
\label{evaluation_setup}

We evaluate \GALA by addressing the following research questions:

\begin{itemize}[leftmargin=*]
    \item \textbf{RQ1:} How effective is \GALA in root cause ranking compared to state-of-the-art methods?
    \item \textbf{RQ2:} How effective is \GALA in incident summarization and action recommendation?
    \item \textbf{RQ3:} How does each component of \GALA affect overall RCA performance?
    \item \textbf{RQ4:} How does the choice of LLM affect the performance, efficiency, and cost of \GALA?
    \item \textbf{RQ5:} How do parameters in agentic investigation (Phase III) affect \GALA's performance?

\end{itemize}

\subsection{Experiment Setup}

\stitle{Microservice Applications}
We evaluate \GALA on two microservice benchmark datasets.
The \textit{OnlineBoutique} (OB) dataset comprises 90 injected-fault scenarios across five services of a 17-service e-commerce application communicating via RPC, spanning six fault types (CPU hog, memory leak, disk I/O stress, network delay, socket errors, and packet loss) with three independent repetitions each~\cite{pham2025benchmark}.
The \textit{TrainTicket} (TT) dataset is a larger and more complex benchmark based on a ticket-booking system with 69 microservices, also comprising 90 fault scenarios across the same six fault types~\cite{pham2025benchmark}.
Together, these two datasets provide complementary evaluation environments: OB tests precision on a compact, well-connected service graph, while TT stresses scalability and reasoning depth on a significantly larger topology.

We additionally provide results on the \textit{AegisLab} benchmark~\cite{aegislab}, which spans 25 distinct fault types, using 100 sampled fault cases. This additional evaluation compares \GALA against a representative subset of baselines; detailed results are reported in our technical report~\cite{gala_technical_report}, with the main findings summarized in Section~\ref{sec:RQ1}.

\stitle{LLM Settings}
We evaluate \GALA with seven LLMs spanning both proprietary and open-source families. Proprietary models are accessed via their official APIs: \textit{gemini-2.5-flash}~\cite{gemini_2_5_flash_2025}, \textit{claude-sonnet-4-5}~\cite{claude_sonnet_4_5_2025}, \textit{GPT-4.1-mini}~\cite{openai_gpt41_mini_2025}, and \textit{GPT-4.1}~\cite{openai_gpt41_2025}. Open-source models---\textit{deepseek-v3.1}~\cite{DeepSeek-V3_1_2025}, \textit{Qwen3-Next-80B-A3B-Instruct}~\cite{Qwen3_80B_2025}, and \textit{llama-4-maverick}~\cite{Llama-4-Maverick_2025}---are served via the Together\,AI inference platform~\cite{together-ai-pricing}. The selection of both large and small models is informed by Company~A's real-world AIOps deployment experience: in production-grade incident management pipelines, thousands of incidents are analyzed daily, making cost-effectiveness a first-class requirement alongside diagnostic accuracy. Accordingly, we designate \textit{GPT-4.1-mini} as the default backbone for all reported \GALA results, as it delivers near-frontier accuracy at a fraction of the cost of larger models (Table~\ref{tab:llm_comparison}). All models are invoked with temperature 1.0.

\stitle{Parameter Settings} For the Graph-Guided Investigation Algorithm, the default settings in \GALA are: parallel branches $k{=}6$, pruning threshold $\theta{=}0.6$, batch size $b{=}2$.

\stitle{Baselines} We compare \GALA against two categories of baselines. \textit{Non-LLM baselines}: Granger~\cite{lin2024root}, CausalRCA~\cite{xin2023causalrca}, and BARO~\cite{pham2024baro}, representing RCA paradigms based on temporal dependency analysis, structured causal modeling, and graph-centrality reasoning, respectively. We also report results for PC~\cite{spirtes2001causation}, LiNGAM~\cite{shimizu2006lingam}, TraceRCA~\cite{li2021practical}, CIRCA~\cite{li2022causal}, RCD~\cite{rcd}, HeMiRCA~\cite{zhu2024hemirca}, and Nezha~\cite{yu2023nezha}. For methods requiring causal graphs for root-cause ranking, we distinguish PageRank-based~\cite{bianchini2005inside} (-PR) and random-walk-based~\cite{lawler2010random} (-RW) variants with corresponding suffixes.

\textit{LLM-based baselines}: We include mABC~\cite{zhangMABCMultiAgentBlockchaininspired2024}, Flow-of-Action~\cite{pei2025flow},  RCA-Agent from OpenRCA~\cite{xuopenrca}, and \GALAArxiv~\cite{tian2025gala_arxiv}, a previous version of this work that differs from the current \GALA in several design components. These represent recent multi-agent and agentic LLM approaches to RCA. We further implement two additional LLM-based baselines to broaden the comparison. For GraphRAG~\cite{edge2024local}, we augment the service dependency graph $\GD$ with nodes representing service metrics and logs. For ReAct~\cite{yao2023react}, we implement an agent that iteratively reasons over telemetry with a fixed action space covering service inspection, neighbor traversal, metric and log queries, and final ranking. All baselines except Nezha have released source code; we reimplemented Nezha on a best-effort basis.

\stitle{Environment Details}
All experiments were executed on an Ubuntu 22.04 virtual machine with eight physical CPU cores, 16\,GB RAM, and 500\,GB SSD storage, using Python 3.10.12 without GPU.

\subsection{Evaluation Metrics}
\stitle{Ranking metrics (RQ1, RQ3--5)}
Following standard practice in the RCA literature~\cite{pham2025benchmark,wu2020microrca,lin2024root,pham2024baro,yu2023nezha,pham2024root}, we report three ranking metrics: \textit{AC@1}, the proportion of incidents for which the top-ranked candidate is the true root cause; \textit{AC@3}, the proportion of incidents for which the true root cause appears in the top three; and \textit{MRR}, the mean reciprocal rank of the true root cause across all incidents.
All three metrics are computed at the pod level: a ranked candidate counts as correct when it matches the ground-truth faulty pod, and no fault-type match is required.

\stitle{Text quality metrics (RQ2)}
We measure \GALA's summarization and recommendation quality using BERTScore~\cite{zhang2020bertscore} (BE), cosine similarity~\cite{salton1986introduction} (COS),  ROUGE-L~\cite{lin2004rouge} (RL), and METEOR~\cite{banerjee2005meteor} (MT) as traditional NLG baselines, and our \surescore (Section~\ref{SURE_Score}) as the primary evaluation framework.

\subsection{Evaluation Results}

\subsubsection{RQ1: Overall Root Cause Ranking Effectiveness.}
 
\label{sec:RQ1}

\begin{table*}[h]
\centering
\renewcommand{\arraystretch}{0.85}
\caption{RQ1 results on Online Boutique (OB) and Train Ticket (TT). Best in \textbf{bold}, second best \underline{underlined}. Improvement =\,relative ranking accuracy improvement over \underline{second best} in column. }
    \vspace{-3mm}
\label{tab:main_results}
\resizebox{\textwidth}{!}{%
\begin{tabular}{ll|ccccc|ccccc}
\toprule
& & \multicolumn{5}{c|}{\textbf{Online Boutique}} & \multicolumn{5}{c}{\textbf{Train Ticket}} \\
\cmidrule(lr){3-7} \cmidrule(lr){8-12}
Type & Method & AC@1 (\%) & AC@3 (\%) & MRR & Time~(s) &Token$_{\text{total}}$ & AC@1 (\%) & AC@3 (\%) & MRR & Time~(s) & Token$_{\text{total}}$ \\
\midrule
\multirow{13}{*}{\rotatebox[origin=c]{0}{Non-LLM}}
 & BARO               & 14.44 & 87.78 & 0.524 & \textbf{0.05}  & -      & \underline{66.67} & \underline{82.22} & \underline{0.759} & \textbf{0.27}  & - \\
 & CausalRCA          & 26.67 & 68.89 & 0.490 & 1{,}575  & -   & -     & -     & -     & T/O    & - \\
 & PC-PR              & 15.56 & 36.67 & 0.332 & 6.68     & -   & 1.11  & 6.67  & 0.082 & 1{,}746  & - \\
 & PC-RW              & 3.33  & 43.33 & 0.290 & 6.57     & -   & 1.11  & 4.44  & 0.077 & 1{,}799  & - \\
 & LiNGAM-PR          & 0.00  & 1.11  & 0.086 & 0.50     & -   & 0.00  & 0.00  & 0.021 & 48.90  & - \\
 & LiNGAM-RW          & 3.33  & 43.33 & 0.292 & 0.69     & -   & 1.11  & 4.44  & 0.079 & 60.53  & - \\
 & Granger-PR         & 12.22 & 35.56 & 0.313 & 10.86    & -   & 3.33  & 6.67  & 0.087 & 400    & - \\
 & Granger-RW         & 3.33  & 44.44 & 0.293 & 11.15    & -   & 1.11  & 3.33  & 0.076 & 406    & - \\
 & TraceRCA           & 13.19 & 61.54 & 0.340 & 3.23     & -   & 64.44 & 75.56 & 0.711 & 13.45  & - \\
 & Nezha              & 28.89 & 52.22 & 0.436 & 76.21    & -   & 15.56 & 40.00 & 0.324 & 513    & - \\
 & CIRCA     & \underline{66.67} & \underline{90.00} & \underline{0.789} & 5.10  & - & 53.33 & 71.11 & 0.636 & 2{,}164 & - \\
 & HeMiRCA   & 37.78 & 58.89 & 0.503 & \underline{0.10} & - & 21.11 & 34.44 & 0.322 & \underline{0.40} & - \\
 & RCD       & 64.44 & 72.22 & 0.713 & 51.90 & - & 10.00 & 18.89 & 0.151 & 164.9 & - \\
\midrule
\multirow{6}{*}{\rotatebox[origin=c]{0}{LLM-based}}
 & GraphRAG           & 21.35 & 35.96 & 0.331 & 9.58   & 78.2K    & 7.78  & 14.44 & 0.163 & 19.62  & 149.3K \\
 & ReAct              & 44.44 & 65.56 & 0.549 & 14.73  & \underline{12.6K}    & 24.44 & 41.39 & 0.353 & 15.92  & \underline{28.3K} \\
 \cmidrule(l){2-12}
 & mABC               & 13.33 & 44.44 & 0.296 & 57.02  & 21.9K    & 12.22 & 23.33 & 0.196 & 92.98  & 38.9K \\
 & RCA-Agent          & 20.00 & 48.89 & 0.337 & 50.63  & 60.0K    & 32.22 & 58.89 & 0.440 & 89.81  & 116.4K \\
 & \GALAArxiv &  38.89  & 86.67  & 0.639  & 49.48    &  13.8K   & 41.94  & 72.04 & 0.602 &  136.33 & 37.6K \\
 & Flow-of-Action     & 48.35 & 70.33 & 0.590 & 36.50  & \textbf{9.6K}     & 47.78 & 74.44 & 0.627 & 123    & \textbf{24.5K} \\
 \cmidrule(l){2-12}
 & \textbf{GALA+}      & \textbf{74.44} & \textbf{98.89} & \textbf{0.854} & 52.87 & 32.7K  & \textbf{73.33} & \textbf{85.56} & \textbf{0.801} & 63.16  & 90.4K \\
 & Improvement          & +12\% & +10\% & +8\% & - & - & +10\% & +4\% & +6\% & - & - \\
\bottomrule
\end{tabular}%
}
\end{table*}

Table~\ref{tab:main_results} reports accuracy, runtime, and total token consumption for all methods on OB and TT. \GALA achieves the highest performance across all ranking metrics on both datasets, attaining {74.44\%} AC@1 and {98.89\%} AC@3 on OB, and {73.33\%} AC@1 and {85.56\%} AC@3 on TT.
 
\stitle{Comparison with Non-LLM Baselines}
The strongest non-LLM baseline differs by system: CIRCA leads on OB (66.67\% AC@1, 90.00\% AC@3, and 0.789 MRR), whereas BARO leads on TT (66.67\% AC@1, 82.22\% AC@3, and 0.759 MRR). \GALA outperforms the best non-LLM method on each system, by +7.77 and +6.66 AC@1 percentage points on OB and TT, respectively.
On OB, BARO frequently overpredicts Redis because its disk I/O exhibits persistent spikes across many incidents, causing this spurious signal to dominate the ranking. This failure mode highlights the risk of relying on a single telemetry modality for RCA. CausalRCA times out entirely on TT (over one hour per case), highlighting the scalability limitations of causal structure learning on larger service graphs.
 
\stitle{Comparison with LLM-based Baselines}
Flow-of-Action is the strongest prior LLM-based approach (48.35\% AC@1 on OB; 47.78\% on TT), yet \GALA surpasses it by {+26.09} and {+25.55} percentage points in AC@1, respectively, showing the advantage of graph-guided investigation over fixed procedural flows. GALA is outperformed by \GALA because it refines its root cause hypothesis from a single retrieved telemetry bundle for one pod at a time, which often results in insufficient evidence gathering and validation. mABC performs poorly on both datasets because its blockchain-inspired collaboration protocol frequently hallucinates non-existent service names. GraphRAG consumes the most tokens among all methods (78.2K on OB; 149.3K on TT) yet achieves only 21.35\% and 7.78\% AC@1, confirming that graph-based retrieval without iterative multi-modal reasoning is insufficient for precise fault localization. ReAct improves over GraphRAG while using far fewer tokens, but its unconstrained action space often causes the investigation to drift away from plausible root-cause candidates, especially on TT's larger service graph.

\stitle{Time and Token Efficiency} \GALA requires about one minute per incident on average, comparable to other LLM-based methods such as RCA-Agent and Flow-of-Action. Although non-LLM methods such as BARO are faster, \GALA is substantially more robust in identifying the true root cause and additionally produces rich, actionable textual outputs, making the added runtime acceptable in practice. Notably, CIRCA, RCD, PC, and Granger are faster than \GALA on OB but scale poorly on TT, becoming much slower as the dependency graph expands.

\GALA consumes 32.7K total tokens on OB and 90.4K on TT, reflecting the cost of exploring multiple candidate branches and retaining reasoning traces for final report generation. Even so, it uses roughly half the tokens of RCA-Agent on OB (32.7K vs.\ 60.0K) while achieving 3.7$\times$ higher AC@1 and lower runtime (52.87s vs.\ 89.81s), as graph-guided pruning avoids exhaustive tool-call iteration. Flow-of-Action uses fewer tokens, but at the cost of roughly 26 percentage points in AC@1. Overall, \GALA's token overhead is a deliberate trade-off for substantially stronger diagnostic performance.

\stitle{Statistical Significance}
We apply McNemar's test~\cite{pembury2020effective} on per-incident majority-vote outcomes across three runs to assess whether \GALA's improvement over Flow-of-Action is statistically significant. On OB, \GALA correctly diagnoses 27 incidents that Flow-of-Action misses versus 4 in the reverse direction; on TT, 35 vs.\ 12. Both asymmetries are significant ($p < 0.001$, one-sided exact binomial).

\stitle{AegisLab Results}
AegisLab covers 25 distinct fault types, compared with six in OB and TT. On 100 sampled cases, \GALA achieves the best MRR, AC@1, and AC@3 among the evaluated methods. It improves MRR over the strongest baseline, Flow-of-Action, from 0.38 to 0.48 (+26\% relative, comparable to its +28\% relative margin on TT), and AC@3 from 0.52 to 0.62. The detailed comparison is reported in our technical report~\cite{gala_technical_report}.

\begin{rqbox}{RQ1 Takeaway}
\GALA achieves the strongest root-cause ranking performance on all datasets, and its improvement over the best prior LLM-based baseline is statistically significant. Although \GALA incurs moderate runtime and token overhead, this cost is offset by substantially higher diagnostic accuracy and actionable output generation.
\end{rqbox}

\subsubsection{RQ2: Incident Summarization and Action Recommendation}
\label{sec:RQ2}

\begin{table}[t]
\centering
\renewcommand{\arraystretch}{0.82}
\caption{RCA text report evaluation. Traditional metrics: BE\,=\,BERTScore, CS\,=\,Cosine Similarity, RL\,=\,ROUGE-L, MT\,=\,METEOR. \surescore results with two LLMs (Q\,=\,\textit{Qwen3}, L\,=\,\textit{Llama-4}), E\,=\,Evidence Groundedness, O\,=\,Operational Feasibility, D\,=\,Diagnostic Precision, T\,=\,Triage Logic, F\,=\,Final (mean). Best in {bold}.}
\label{tab:sure_score}

\vspace{-1mm}

\resizebox{\columnwidth}{!}{%
\addtolength{\tabcolsep}{-2.5pt}

\begin{tabular}{cl|c|c|c|c|c|c|c|c|c|c|c|c|c|c}
\toprule
& & \multicolumn{4}{c|}{\textbf{Trad.}} & \multicolumn{5}{c|}{\textbf{\surescore (Q)}} & \multicolumn{5}{c}{\textbf{\surescore (L)}} \\
\cmidrule(lr){3-6} \cmidrule(lr){7-11} \cmidrule(lr){12-16}
 & Method & BE & CS & RL & MT & E & O & D & T & \textbf{F} & E & O & D & T & \textbf{F} \\
\midrule
\multirow{5}{*}{\rotatebox[origin=c]{90}{OB}}
 & BARO+LLM      & -.10 & .28 & \textbf{.13} & .16 & 1.0 & 1.4 & 1.7 & 1.3 & 1.3 & 1.0 & 3.4 & 2.2 & 2.2 & 2.2 \\
 & mABC          & -.23 & .21 & .10 & .15 & 1.0 & 1.0 & 1.8 & 1.4 & 1.3 & 1.3 & 1.0 & 2.6 & 3.7 & 2.2 \\
 & Flow-of-A.    & -.23 & .19 & .11 & .14 & 1.4 & 1.0 & 2.3 & 1.0 & 1.4 & 1.7 & 1.0 & 2.4 & 1.2 & 1.6 \\
 & RCA-Agent     & -.44 & .08 & .09 & .01 & 1.0 & 1.0 & 1.6 & 1.0 & 1.2 & 1.0 & 1.0 & 1.2 & 1.0 & 1.0 \\
 & \textbf{GALA+} & \textbf{.04} & \textbf{.35} & \textbf{.13} & \textbf{.21} & \textbf{1.8} & \textbf{3.5} & \textbf{4.4} & \textbf{3.3} & \textbf{3.3} & \textbf{5.0} & \textbf{3.7} & \textbf{4.5} & \textbf{5.0} & \textbf{4.5} \\
\midrule
\multirow{5}{*}{\rotatebox[origin=c]{90}{TT}}
 & BARO+LLM      & -.06 & .37 & \textbf{.17} & .15 & 1.0 & 1.4 & 2.1 & 1.4 & 1.5 & 1.0 & 3.1 & 2.5 & 1.9 & 2.1 \\
 & mABC          & -.17 & .25 & .11 & .16 & 1.0 & 1.0 & 2.1 & 1.4 & 1.4 & 1.2 & 1.0 & 2.8 & 3.1 & 2.0 \\
 & Flow-of-A.    & -.25 & .23 & .13 & .14 & 1.3 & 1.0 & 2.8 & 1.0 & 1.5 & 2.3 & 1.0 & 2.8 & 1.2 & 1.8 \\
 & RCA-Agent     & -.41 & .14 & .13 & .01 & 1.0 & 1.0 & 1.8 & 1.0 & 1.2 & 1.0 & 1.0 & 1.2 & 1.0 & 1.1 \\
 & \textbf{GALA+} & \textbf{.06} & \textbf{.38} & .13 & \textbf{.21} & \textbf{1.4} & \textbf{3.3} & \textbf{4.2} & \textbf{3.2} & \textbf{3.0} & \textbf{4.9} & \textbf{3.5} & \textbf{4.4} & \textbf{4.8} & \textbf{4.4} \\
\bottomrule
\end{tabular}%
}
\end{table}

Table~\ref{tab:sure_score} reports \surescore results evaluated by two independent LLMs from different model families---\textit{Qwen3} and \textit{Llama-4-Maverick}---alongside traditional natural language generation metrics. Since BARO produces only fault rankings, we evaluate BARO+LLM, which uses the same backbone (\textit{GPT-4.1-mini}) as \GALA for fair comparison.

\stitle{Traditional Metrics are insufficient}
\GALA attains the best BERTScore, cosine similarity, and METEOR on both OB and TT, yet all four metrics are weakly discriminative: BERTScore is mostly negative, while cosine similarity and ROUGE-L vary little across methods.
For example, on TT, BARO+LLM attains nearly the same cosine similarity as \GALA (0.37 vs.\ 0.38) but a much lower \surescore\ (2.1 vs.\ 4.4), showing that surface-level text similarity fails to capture the causal grounding and operational specificity required for RCA reports.

\stitle{\surescore results}
Across both evaluator LLMs, \GALA consistently achieves the highest overall \surescore\ on both OB and TT, and ranks first on all evaluation dimensions. Under Qwen3, \GALA scores 3.3 on OB and 3.0 on TT; under \textit{Llama-4-Maverick}, it scores 4.5 on OB and 4.4 on TT. Although the absolute scores differ and \textit{Llama-4-Maverick} tends to give a higher score, the two evaluators agree on the central result that \GALA clearly outperforms all baselines, with only minor variation in the ranking among baselines. This overall consistency in ranking suggests that \surescore\ captures robust quality differences rather than evaluator-specific artifacts. The largest gains appear in Operational Feasibility and Diagnostic Precision, reflecting \GALA’s ability to translate accurate diagnoses into specific, implementable recommendations.

\stitle{Evidence Groundedness Diverges Across Evaluators}
This dimension shows the greatest divergence between evaluators. Under \textit{Qwen3}, all methods receive scores close to 1.0, including \GALA (1.8 on OB), whereas under \textit{Llama-4-Maverick}, \GALA reaches 5.0 while the baselines remain low. This contrast suggests that the two evaluators apply different standards for explicit evidential support: \textit{Qwen3} favors verbatim citation of logs or metric values, whereas \textit{Llama-4-Maverick} also credits paraphrased textual references. Despite this difference in grading strictness, \GALA ranks highest under both evaluators, indicating that its multi-modal reasoning consistently yields more evidence-grounded narratives.

\begin{table}[t]
\centering
\renewcommand{\arraystretch}{0.85}
\vspace{-3mm}
\caption{Human SRE evaluation. E/O/D/T/F as in Table~\ref{tab:sure_score}.}
\label{tab:human_eval}
\vspace{-2mm}
\resizebox{0.68\columnwidth}{!}{%

\begin{tabular}{ll|ccccc}
\toprule
 & Method & E & O & D & T & \textbf{F} \\
\midrule
\multirow{2}{*}{OB}
& BARO+LLM      & 2.3 & 2.1 & 2.3 & 2.8 & 2.4 \\
 & Flow-of-Action & 3.1 & 1.4 & 3.4 & 2.4 & 2.6 \\
 & \textbf{GALA+}  & \textbf{4.2} & \textbf{4.4} & \textbf{4.6} & \textbf{4.5} & \textbf{4.4} \\
\midrule
\multirow{2}{*}{TT}
 & BARO+LLM       & 2.3 & 2.0 & 2.4 & 2.8 & 2.4 \\
 & Flow-of-Action & 2.8 & 1.5 & 3.3 & 2.5 & 2.6 \\
 & \textbf{GALA+}  & \textbf{3.8} & \textbf{4.6} & \textbf{4.4} & \textbf{4.0} & \textbf{4.2} \\
\bottomrule
\end{tabular}%
}
\vspace{-3mm}
\end{table}

\stitle{Human Evaluation}
To validate that \surescore aligns with practitioner judgment, three SREs with production incident response experience independently graded 540 reports (90~incidents $\times$ 3~methods $\times$ 2~datasets) on the same four dimensions using the same 1--5 scale. Table~\ref{tab:human_eval} reports the averaged results. \GALA outperforms Flow-of-Action across all dimensions, with the largest gap on Operational Feasibility (+3.0 average): Flow-of-Action produces no remediation recommendations and scores near the floor (1.4--1.5), while \GALA's three-tier stratified recommendations consistently exceed 4.0. 
The BARO+LLM's overall performance is low because its narratives do not cite any concrete telemetry.
The evaluators independently identified Evidence Groundedness as a weak dimension for all methods, confirming that LLMs paraphrase rather than cite telemetry and validating that \surescore captures a genuine quality gap. Human scores (2.4--4.4) are directionally consistent with automated \surescore ratings, though higher in absolute terms because evaluators give partial credit for implicit evidence references. This offset does not affect method ranking: both automated and human evaluations agree that \GALA dominates on every dimension.

\stitle{Effect of Report Length}
While \surescore does not normalize for report length, it is not driven by length either: the checklist requires exact service and pod identifiers, quantitative impact figures, evidence-anchored reasoning steps, and implementable remediation actions, rather than generic long-form descriptions. 
On OB, mABC produces \emph{longer reports than \GALA on average (339 vs.\ 322 tokens) yet is graded 2.2 against \GALA's 4.5 by \textit{Llama-4-Maverick} and 1.3 against 3.3 by \textit{Qwen3}. This suggests that \surescore is driven more by diagnostic precision and operational usefulness than by length alone.}

\stitle{Evaluator Agreement}
To quantify \surescore reliability, we assess two forms of agreement. First, pooled inter-evaluator agreement between \textit{Qwen3} and \textit{Llama-4-Maverick} across all methods, dimensions, and datasets yields Kendall's $\tau = 0.68$ ~\cite{sen1968estimates} and Spearman's $\rho = 0.84$ \cite{schmid2007multivariate} (both $p < 0.001$, $n = 50$), confirming that the two independent evaluator LLMs produce consistent method rankings despite differing in absolute scoring scales. Second, Spearman correlation between averaged automated \surescore and the independent human SRE ratings (Table~3) is $\rho = 0.92$ ($p < 0.001$, $n = 20$), indicating that \surescore serves as a reliable proxy for expert assessment.

\begin{rqbox}{RQ2 Takeaway}
Traditional NLG metrics fail to differentiate RCA report quality. \GALA outperforms all baselines under \surescore, confirmed by cross-model evaluator agreement and independent human SRE evaluation. Evidence Groundedness remains an open challenge.
\end{rqbox}

\subsubsection{RQ3: Ablation Study}

\begin{figure}[h]
    \centering
    \centerline{
        \includegraphics[width=0.495\textwidth]{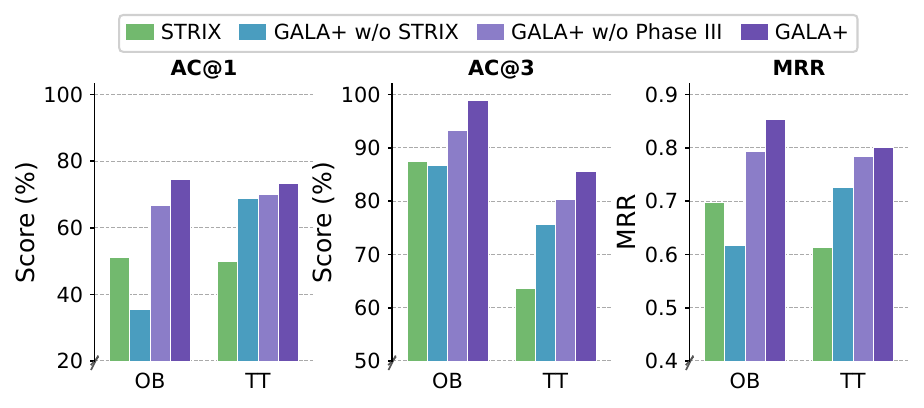}
    }
    \vspace{-4mm}
    \caption{Ablation study of \GALA.} 
    \vspace{-4mm}
    \label{fig:gala_ablation}
\end{figure}

\begin{table}[h]
\centering
\renewcommand{\arraystretch}{0.85}

\caption{Ablation study of \STRIX\ components.}
\vspace{-3mm}
\label{tab:strix_ablation}

\resizebox{0.81\columnwidth}{!}{
\begin{tabular}{clccc}
\toprule
Dataset & Method & AC@1 (\%) & AC@3 (\%) & MRR \\
\midrule
\multirow{4}{*}{OB}
 & \STRIX w/o Instability  & 18.18 & 78.41 & 0.449 \\
 & \STRIX w/o Centrality   & 44.32 & 75.00 & 0.628 \\
 & \STRIX w/o Impact       & 36.36 & 69.32 & 0.559 \\
 & \STRIX                  & \textbf{51.14} & \textbf{87.50} & \textbf{0.697} \\
\midrule
\multirow{4}{*}{TT}
 & \STRIX w/o Instability  & 9.09  & 39.77 & 0.307 \\
 & \STRIX w/o Centrality   & 26.14 & 68.18 & 0.498 \\
 & \STRIX w/o Impact       & 19.32 & \textbf{69.32} & 0.487 \\
 & \STRIX                  & \textbf{50.00} & 63.64 & \textbf{0.613} \\
\bottomrule
\end{tabular}
}
\vspace{-3mm}
\end{table}

Figure~\ref{fig:gala_ablation} presents an ablation study isolating the contribution of \GALA's core components, and Table~\ref{tab:strix_ablation} further ablates \STRIX's three structural scoring dimensions.

\stitle{GALA+ Component Ablation}
Compared to full \GALA, using \STRIX alone reduces AC@1 to 51.14\% on OB and 50.00\% on TT, showing that trace-based structural scoring is informative but not sufficient on its own. Removing \STRIX causes a substantial drop on OB (from 74.44\% to 35.56\%) and a smaller drop on TT (from 73.33\% to 68.89\%), indicating that its contribution is dataset-dependent. In particular, \STRIX is especially valuable on OB, where metric-only signals are more easily dominated by noisy downstream effects and thus benefit more from trace- and structure-aware ranking. Removing Phase III (Section \ref{LLM Agentic Reasoning and Re‐ranking}) also degrades performance, showing that graph-guided agentic investigation consistently refines the initial ranking by re-examining top candidates through localized multi-modal evidence. 

\stitle{Agent-Level Ablation}
The ablations above remove entire phases. To test whether \GALA's LLM agents add value beyond deterministic aggregation over the same inputs, we replace each agent with a fixed rule on OB, leaving the rest of the pipeline unchanged. Substituting reciprocal-rank fusion for the \textsf{Consolidation Agent lowers AC@1 from 74.44\% to 63.33\%, AC@3 from 98.89\% to 96.70\%, and MRR from 0.854 to 0.790. When the \textsf{Evidence Synthesis Agent} is replaced with threshold-count aggregation over \textsc{DeepDive} results in full \GALA, performance similarly drops to 0.790 MRR and 63.33\% AC@1. These results indicate that both the \textsf{Consolidation Agent} and \textsf{Evidence Synthesis Agent} contribute beyond simple aggregation rules. }

\stitle{\STRIX Dimension Ablation}
Among the three dimensions, Instability ($\phi_1$) is the most discriminative: removing it drops AC@1 from 51.14\% to 18.18\% on OB and from 50.00\% to 9.09\% on TT. Removing Centrality ($\phi_2$) or Impact ($\phi_3$) causes smaller but still clear degradations, confirming that both provide complementary structural and performance signals beyond tail-latency instability alone. On TT, \STRIX\ without Centrality or Impact still attains relatively high AC@3 (68.18\% and 69.32\%), suggesting that Instability is already effective at retaining the true root cause within the top candidates, while $\phi_2$ and $\phi_3$ are especially important for sharpening the final top-1 ranking.

\begin{rqbox}{RQ3 Takeaway}
Each major component of \GALA contributes meaningfully to RCA performance: \STRIX supplies complementary trace- and structure-aware ranking signals, and graph-guided agentic investigation further sharpens the already strong initial hypotheses.
\end{rqbox}

\subsubsection{RQ4: Impact of LLM Selection}
\label{sec:RQ4}

\begin{table}[t]
\centering
\renewcommand{\arraystretch}{0.85}

\caption{Impact of LLM choice on \GALA (OB dataset). AC in~\%, Token\,=\,total tokens, \$\,=\,cost per incident. Best in \textbf{bold}.}
\label{tab:llm_comparison}
\vspace{-2mm}

\resizebox{0.9\columnwidth}{!}{%
\addtolength{\tabcolsep}{-1.5pt}    

\begin{tabular}{lcccccc}
\toprule
LLM & AC@1 & AC@3 & MRR & Token & Time (s) & \$ \\
\midrule
\textit{gemini-2.5-flash}  & 55.6 & 87.8 & .718 & \textbf{17.8K} & \textbf{18.5}  & .030 \\
\textit{deepseek-v3}       & 67.8 & 94.4 & .810 & 35.2K & 64.9  & .044 \\
\textit{claude-sonnet-4-5} & 71.1 & 93.3 & .827 & 45.6K & 101.3 & .255 \\
\textit{GPT-4.1-mini}      & 74.4 & \textbf{98.9} & .854 & 32.7K & 52.9  & .022 \\
\textit{Qwen3-80B}         & 76.7 & 95.6 & .855 & 40.2K & 21.3  & .015 \\
\textit{llama-4-maverick}  & 77.8 & 97.8 & .872 & 32.0K & 37.4  & \textbf{.003} \\
\textit{gpt-4.1}           & \textbf{78.9} & 94.4 & \textbf{.873} & 34.0K & 48.4  & .110 \\
\bottomrule
\end{tabular}%
}
\vspace{-3mm}
\end{table}

Table~\ref{tab:llm_comparison} reports accuracy, total token consumption, latency, and cost per incident for seven LLMs on OB. AC@1 broadly increases with model capability, from \textit{gemini-2.5-flash} (55.6\%) to \textit{gpt-4.1} (78.9\%). Two open-source models---\textit{llama-4-maverick} (77.8\%) and \textit{Qwen3-80B} (76.7\%)---approach top-tier accuracy at a fraction of the cost (\$0.003 and \$0.015 per incident via Together\,AI), with \textit{llama-4-maverick} offering the best overall trade-off: near-top accuracy at the lowest cost among all evaluated models. We designate \textit{GPT-4.1-mini} as the default backbone for reproducibility via its official API, noting that it still achieves 74.4\% AC@1 at \$0.022---only 4.5 points below \textit{gpt-4.1} at one-fifth the cost. Wall-clock time varies considerably (18.5s--101.3s), driven primarily by API response latency rather than token volume; practitioners with strict latency constraints may prefer \textit{Qwen3-80B} (21.3s) or \textit{gemini-2.5-flash} (18.5s).

\begin{rqbox}{RQ4 Takeaway}
\GALA benefits from stronger LLMs, but open-source models can deliver near-top performance at substantially lower cost.
\end{rqbox}

\subsubsection{RQ5: Impact of Agentic Workflow Parameters}

\label{sec:RQ5}

\begin{figure}[h]
    \centering
    \vspace{-4mm}
    \centerline{
        \includegraphics[width=0.5\textwidth]{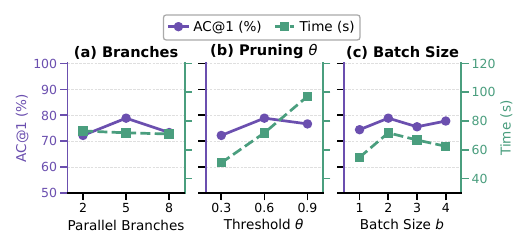}
    }
    \vspace{-4mm}
    \caption{GALA+ parameter study results.} 
    \label{fig:gala_parameter}
\end{figure}

Figure~\ref{fig:gala_parameter} reports AC@1 and execution time of \GALA on TT with different combinations of the three key agentic parameters: parallel branches $k$, pruning threshold $\theta$, and batch size $b$.

\noindent\textbf{Parallel branches ($k$).}
A moderate value of $k$ yields the best accuracy, whereas both smaller and larger settings perform worse. This suggests that investigating too few initial candidates limits coverage, while too many introduce noisier hypotheses without improving the final ranking. Execution time remains largely unchanged across settings because these investigation branches are launched in parallel.

\noindent\textbf{Pruning threshold ($\theta$).}
Performance peaks around $\theta{=}0.6$. Lower confidence thresholds cause agents to terminate too early, before following potentially informative causal chains, whereas higher thresholds encourage unnecessary graph expansion and slightly reduce accuracy while increasing latency.

\noindent\textbf{Batch size ($b$).}
A small batch size works best, with performance peaking around $b{=}2$. Smaller batches may limit local coverage, whereas larger batches introduce noisier neighbors and reduce the focus of investigation. Execution time varies only modestly across settings.

\noindent\textbf{Neighbor selection.}
\GALA samples a batch of $b$ neighbors uniformly at random during agentic graph-guided investigation. We compare this against an evidence-driven alternative that orders the neighbors of a service by their \STRIX ranking and expands the top $b$. On a 30-case OB subset with five seeds, random sampling attains AC@1 $0.76 \pm 0.03$, AC@3 $0.97 \pm 0.02$, and MRR $0.87 \pm 0.02$, against $0.72 \pm 0.04$, $0.98 \pm 0.02$, and $0.85 \pm 0.02$ for \STRIX-ordered expansion. The \STRIX-based expansion did not dominate random sampling because \STRIX is already used during initial hypothesis generation. Reusing \STRIX again for graph expansion can reinforce the same trace-based bias, whereas random sampling provides more diverse local exploration around the top candidates.

\begin{rqbox}{RQ5 Takeaway}
\GALA is robust across a broad range of parameter settings, and the default configuration provides a strong balance between candidate coverage and inference efficiency.
\end{rqbox}

\section{Case Study}
\label{case_study}
\begin{figure}[h]
    \centering
    \vspace{-1mm}
    \centerline{
        \includegraphics[width=0.46\textwidth]{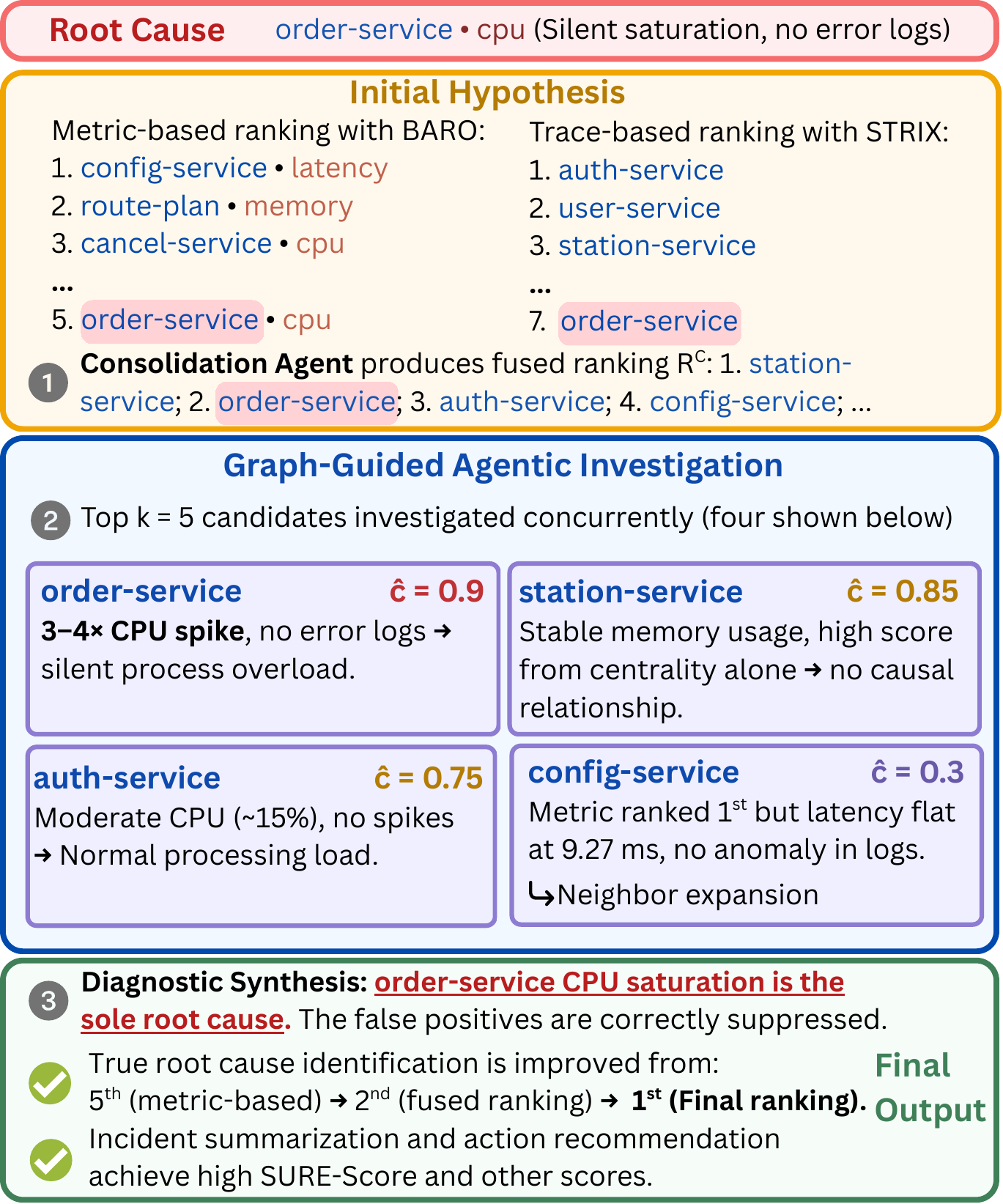}
    }
    \vspace{-1mm}
    \caption{\GALA's reasoning chain for a CPU hog fault case in \texttt{order-service}, ranked top-1 with optimal \surescore.} 
    \vspace{-2mm}
    \label{fig:case_study}
\end{figure}

We illustrate \GALA on a CPU-hog fault occurring in \texttt{order-service} of TT (Figure~\ref{fig:case_study}). The coarse metric-based ranking places \texttt{order-service} at 5th, while \STRIX ranks it 7th based on trace- and structure-aware signals. Although neither modality alone places the true root cause at the top, the fused initial ranking $R^{\text{C}}$ already promotes \texttt{order-service} to 2nd, indicating that Phase I identifies a high-quality set of plausible hypotheses. Phase III then resolves the remaining ambiguity through graph-guided investigation. The \texttt{order-service} branch finds the clearest evidence of the true fault, namely a 3--4$\times$ CPU spike with no error logs, consistent with silent saturation. By contrast, other high-ranked candidates are ruled out through localized multi-modal inspection: their signals are either stable, weak, or unsupported by a causal explanation. After graph-guided investigation and evidence synthesis, \GALA correctly promotes \texttt{order-service} to rank 1 and suppresses the false positives. This case highlights how graph-guided investigation refines an already strong initial ranking into the correct final diagnosis.

\section{Related Work} 

\stitle{LLM for RCA}
Language-model-based RCA falls into three paradigms. Fine-tuning approaches~\cite{saha2022mining,ahmed2023recommending,chen2024automatic,pang2024large,pang2024hybrid,xing2023fusion} adapt pre-trained models to RCA-specific incident profiles. Embedding-based methods~\cite{zhuang2022gan,saha2022mining,chakraborty2023esro,he2023construction,kuang2024knowledge} use vector representations from models like BERT~\cite{devlin2019bert} to capture semantic information in telemetry. Prompt-based methods apply in-context learning~\cite{zhang2024automated,roy2024exploring}, chain-of-thought reasoning~\cite{wei2022chain,ahmed2023recommending,chen2024automatic,zhang2024lm,pei2025flow,xuopenrca}, and retrieval-augmented generation~\cite{lewis2020retrieval,wang2024large,jiang2024xpert,li2024realtcd,li2025coca} to guide reasoning over heterogeneous evidence.
Among agentic approaches, Xu et al.~\cite{xuopenrca} introduce RCA-Agent together with the OpenRCA benchmark, showing both the promise of LLM-based RCA and the substantial challenges that remain even for multi-agent, execution-based reasoning. RCAgent~\cite{wang2024rcagent} achieves strong results through tool-augmented reasoning over logs and code, naturally complementing trace- and metric-driven fault localization rather than targeting it directly. Flow-of-Action~\cite{pei2025flow} encodes standard operating procedures into structured workflows that improve reproducibility on anticipated fault patterns, while naturally emphasizing consistency over flexible exploration in less expected scenarios. mABC~\cite{zhangMABCMultiAgentBlockchaininspired2024} introduces blockchain-inspired multi-agent collaboration for accurate ranking, though our experiments suggest that its decentralized coordination can occasionally lead to less grounded predictions. GALA~\cite{tian2025gala_arxiv} orchestrates a ReAct-style investigation reasoning loop after formulating an initial hypothesis. 
In contrast, \GALA performs graph-guided multi-modal investigation and generates causally grounded incident responses.

\stitle{Non-LLM RCA Methods}
These methods typically rely on causal inference and graph-based analysis. Granger causality~\cite{lin2024root} and structural equation modeling~\cite{xin2023causalrca,shimizu2006lingam,spirtes2001causation} identify relationships in metric time series, while graph-based methods~\cite{pham2024baro,wu2020microrca} exploit service topology through centrality and ranking algorithms~\cite{lawler2010random,bianchini2005inside}. These approaches are often limited to a single modality, make limited use of log semantics, and provide little explanatory support for incident response. More recent multi-modal methods address some of these limitations by fusing heterogeneous observability signals~\cite{tao2024giving,sun2025interpretable,liu2022microcbr}. In particular, Nezha~\cite{yu2023nezha} represents all three modalities as event graphs, MSoFSAnomaly~\cite{fu2025msfsanomaly} selects fault-sensitive features to reduce redundancy, and MULAN~\cite{zheng2024multi} applies contrastive learning for multi-modal fusion.  \GALA preserves each modality in a structured diagnostic bundle and couples statistical hypothesis generation with graph-guided LLM reasoning.

\section{Threats to Validity}
\stitle{Internal Threats} LLM non-determinism at temperature 1.0 introduces output variability; we mitigate this with standardized prompts and report results averaged over three runs. \surescore's checklist was co-developed with experts from a single organization (Company~A), potentially limiting generalizability, though our dual-evaluator and human validation results suggest the dimensions transfer across grading contexts. \GALA requires access to all three telemetry modalities to achieve expected performance.

\stitle{External Threats} Our evaluation uses controlled fault-injection benchmarks with predefined fault types; real-world incidents may exhibit more complex or concurrent failure modes. Prompt templates are designed for the evaluated datasets' telemetry schema; adapting to different deployments requires additional prompt engineering. Token cost scales with the number of candidate services and exploration coverage, which may constrain applicability to very large service graphs.

\section{Conclusions}

This paper presents \GALA, a graph-augmented LLM agentic framework for microservice root cause analysis and incident response generation. By combining dual-signal hypothesis generation via metrics-based ranking and \STRIX with graph-guided investigation over localized multi-modal evidence, \GALA achieves the best overall RCA performance on two benchmarks and outperforms the strongest LLM-based baseline by over 25\% in AC@1; this advantage is statistically significant under McNemar's test ($p < 0.001$). Our \surescore framework, supported by evaluator LLMs and SRE experts, further confirms substantial gains in diagnostic precision and operational feasibility, while identifying evidence groundedness as an open challenge for current LLM-based RCA methods. These results suggest that topology-constrained reasoning is a promising alternative to unconstrained agentic exploration in domains with rich relational structure. Future work will extend \GALA to concurrent multi-fault scenarios and incorporate improved context engineering for stronger evidence grounding.

\section*{Data Availability}
The software underlying this work was developed under a collaborative research agreement with our industrial partner, whose terms preclude public release of the source code. To support reproducibility, the paper reports the complete algorithm specifications, system parameters, workload configurations, and hardware setup used in all experiments. Aggregate measurement data are available from the corresponding author on reasonable request, subject to the partner's approval. The technical report and artifacts permitted for release are available at: \url{https://github.com/EvanTian233/GALA-for-RCA}.

\section*{Acknowledgments}
This research was undertaken, in part, thanks to funding from Bell Canada, NSERC, the Canada Research Chairs Program, and ORF.

\bibliographystyle{ACM-Reference-Format}
\bibliography{gala_plus.bib}

\end{document}